\documentclass[aps,prd,preprintnumbers,showpacs,twocolumn,groupedaddress,nofootinbib]{revtex4}
\usepackage{graphicx}
\usepackage{latexsym}
\def\beq{\begin{equation}}
\def\eeq{\end{equation}}
\def\bey{\begin{eqnarray}}
\def\eey{\end{eqnarray}}

\def\lsim{\mathrel{\raise.3ex\hbox{$<$\kern-.75em\lower1ex\hbox{$\sim$}}}}
\def\gsim{\mathrel{\raise.3ex\hbox{$>$\kern-.75em\lower1ex\hbox{$\sim$}}}}

\begin{document}

\title{Closing Supersymmetric Resonance Regions With Direct Detection Experiments}  
\author{Dan Hooper$^{1,2}$, Chris Kelso$^{3}$, Pearl Sandick$^3$, and Wei Xue$^4$}
\affiliation{$^1$Center for Particle Astrophysics, Fermi National Accelerator Laboratory, Batavia, IL 60510, USA}
\affiliation{$^2$Department of Astronomy and Astrophysics, University of Chicago, Chicago, IL 60637, USA}
\affiliation{$^3$Department of Physics and Astronomy, University of Utah, Salt Lake City, Utah 84112, USA}
\affiliation{$^4$INFN, Sezione di Trieste, SISSA, via Bonomea 265, 34136 Trieste, Italy}

\date{\today}

\begin{abstract}

In order for neutralino dark matter to avoid being overproduced in the early universe, these particles must annihilate (or coannihilate) rather efficiently. Neutralinos with sufficiently large couplings to annihilate at such high a rate (such as those resulting from gaugino-higgsino mixing, as in ``well-tempered'' or ``focus point'' scenarios), however, have become increasingly disfavored by the null results of XENON100 and other direct detection experiments. One of the few remaining ways that neutralinos could potentially evade such constraints is if they annihilate through a resonance, as can occur if 2$m_{\chi^0}$ falls within about $\sim$10\% of either $m_{A/H}$, $m_h$, or $m_Z$. If no signal is observed from upcoming direct detection experiments, the degree to which such a resonance must be tuned will increase significantly. In this paper, we quantify the degree to which such a resonance must be tuned in order to evade current and projected constraints from direct detection experiments. Assuming a future rate of progress among direct detection experiments that is similar to that obtained over the past decade, we project that within 7 years the light Higgs and $Z$ pole regions will be entirely closed, while the remaining parameter space near the $A/H$ resonance will require that $2m_{\chi^0}$ be matched to the central value (near $m_A$) to within less than 4\%. At this rate of progress, it will be a little over a decade before multi-ton direct detection experiments will be able to close the remaining, highly-tuned, regions of the $A/H$ resonance parameter space.

\end{abstract}

\pacs{14.80.Ly, 95.35.+d, 14.80.Da; FERMILAB-PUB-13-077-A}
\maketitle


\section{Introduction}

Much of the motivation for dark matter in the form of weakly interacting massive particles (WIMPs) has been based on the observation that if there exists a stable particle species with a weak-scale mass and annihilation cross section, it will be thermally produced in the early universe in a quantity similar to the measured abundance of dark matter~\cite{KandT}. In no particle physics framework has this been considered in more detail than in that of the minimal supersymmetric Standard Model (MSSM), with dark matter in the form the of the lightest neutralino~\cite{neutralino}. Dark matter candidates motivated by this argument are also interesting from an experimental perspective; it has long been appreciated that such particles could eventually be detected and observed through their interactions with nuclei~\cite{Goodman:1984dc}. As the sensitivity of direct detection experiments has improved, however, this appeal has begun to become something of a liability; for many WIMP candidates, the very interactions that enable them to annihilate and avoid being overproduced in the early universe also lead to elastic scattering cross sections with nuclei that are in excess of the constraints currently provided by the leading direct detection experiments. 

As a simple illustration, consider a dark matter candidate in the form of a Majorana fermion, which annihilates into Standard Model fermions through couplings proportional to mass (such as a neutralino annihilating through the exchange of a pseudoscalar Higgs boson). Turning this annihilation diagram on its side (and replacing the pseuodoscalar with a scalar Higgs boson), this interaction induces an elastic scattering cross section between the dark matter and nuclei, typically dominated by couplings to the strange quark content of the nucleon, and to gluons through heavy quark loops~\cite{Jungman:1995df}. If the exchanged particles are much heavier than the dark matter particle, we can use effective field theory to relate the elastic scattering cross section by a crossing symmetry to the cross section for annihilation~\cite{Beltran:2008xg}. For a 100 GeV dark matter candidate with couplings chosen to yield the desired thermal relic abundance, this leads to a coherent (spin-independent) elastic scattering cross section with nucleons that is on the order of a few times $10^{-7}$ pb. This is in considerable excess of existing constraints from direct detection experiments~\cite{xenon,cdms}. The XENON100 collaboration, in particular, excludes elastic scattering cross sections that are greater than approximately $3 \times 10^{-9}$ pb for 100 GeV WIMPs.

This simple example (and many others like it\footnote{This argument is not limited to Higgs exchange; WIMP candidates capable of coherent (spin-independent) elastic scattering through $Z$-exchange (such as sneutrinos, or fourth generation neutrinos~\cite{sneutrinos}) predict rates at direct detection experiments that are currently ruled out by several orders of magnitude.}) shows that direct detection experiments are currently capable of excluding many dark matter candidates motivated by the thermal production arguments. In light of these considerations, we can ask what types of mechanisms or features might enable a dark matter candidate to evade such constraints. Broadly speaking, there at least five ways in which such arguments can be circumvented:\footnote{We restrict ourselves to WIMP dark matter candidates in this study. Axions, sterile neutrinos, gravitinos and other non-WIMPs could also make up the dark matter without violating constraints from direct detection experiments.}
\begin{itemize}
\item{The dark matter is depleted in the early universe though coannihilations with another particle, rather than through self-annihilations~\cite{Griest:1990kh}.}
\item{Only a subdominant fraction of dark matter annihilations produce quarks, but instead proceed to states such as leptons, gauge bosons, and/or Higgs bosons.}
\item{The WIMP is sufficiently light as to fall below the detection thresholds of the most sensitive direct detection experiments ($m$\,$\lsim$\,$10$ GeV)~\cite{xenon,cdms}.}
\item{After inflation, the universe was never reheated to temperatures sufficient for the dark matter to reach thermal equilibrium.}
\item{The dark matter annihilates efficiently through an $s$-channel resonance~\cite{Griest:1990kh}.}
\end{itemize}

Within the context of the MSSM, three of these possibilities are most commonly discussed. Coannihilation regions of parameter space (with staus, charginos, stops, and other sparticles) are currently viable, so long as the lightest neutralino is only slightly less massive than the next lightest state (typically within a few percent, although stop coannihilations can be effective with larger mass splittings). In the bulk and focus point regions of the MSSM, most neutralino annihilations proceed to gauge and/or Higgs bosons, relaxing the impact of direct detection constraints to some extent. That being said, the most recent constraints from XENON100 appear to rule out nearly all of the remaining focus point parameter space~\cite{Buchmueller:2012hv} (as well as much of the more general parameter space associated with dark matter in the form of a ``well-tempered'' neutralino~\cite{ArkaniHamed:2006mb}); see, however, Ref.~\cite{Feng:2011aa}. If the lightest supersymmetric particle is a relatively pure wino or higgsino, its annihilations can proceed overwhelmingly to gauge boson final states, enabling current direct detection constraints to be evaded, especially if $m_A$ is very large~\cite{Hisano:2012wm}. And lastly, the lightest neutralino can annihilate through the resonant exchange of a Higgs or $Z$ boson. Particularly efficient is the so-called $A$-funnel region of parameter space, in which the resonant exchange of the pseudoscalar Higgs can efficiently deplete the abundance of neutralinos in the early universe without exceeding direct detection constraints.

In this paper, we focus on the last of these possibilities, and determine the level to which the MSSM $A$-funnel resonance must be tuned ({\it ie}. how close to $2 m_{\chi}$ we must set $m_A$) in order to evade the constraints from XENON100 and other direct detection experiments.\footnote{The fine tuning discussed in this paper is distinct from and should not be confused with that associated with the gauge hierarchy problem~\cite{hierarchy}} We also project how this required degree of tuning is expected to increase as direct detection constraints become more stringent (assuming no detection is made), eventually closing the $A$-funnel region entirely. We also present a similar analysis for the light Higgs and $Z$ pole regions, as well as more general arguments pertaining to a broad class of resonant annihilating dark matter models.

\begin{figure*}[t!]
\includegraphics[width=2.9in]{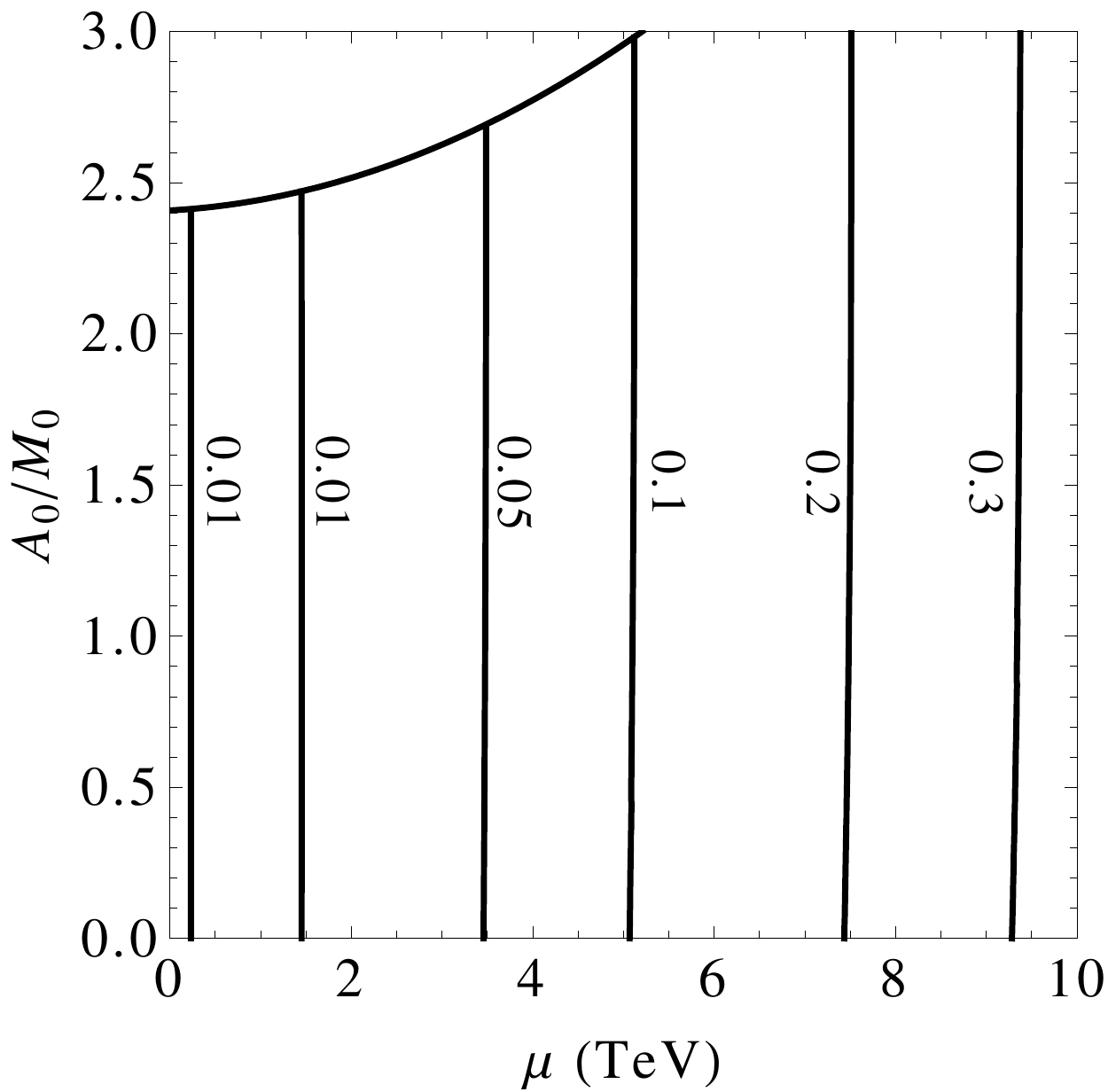}
\includegraphics[width=2.9in]{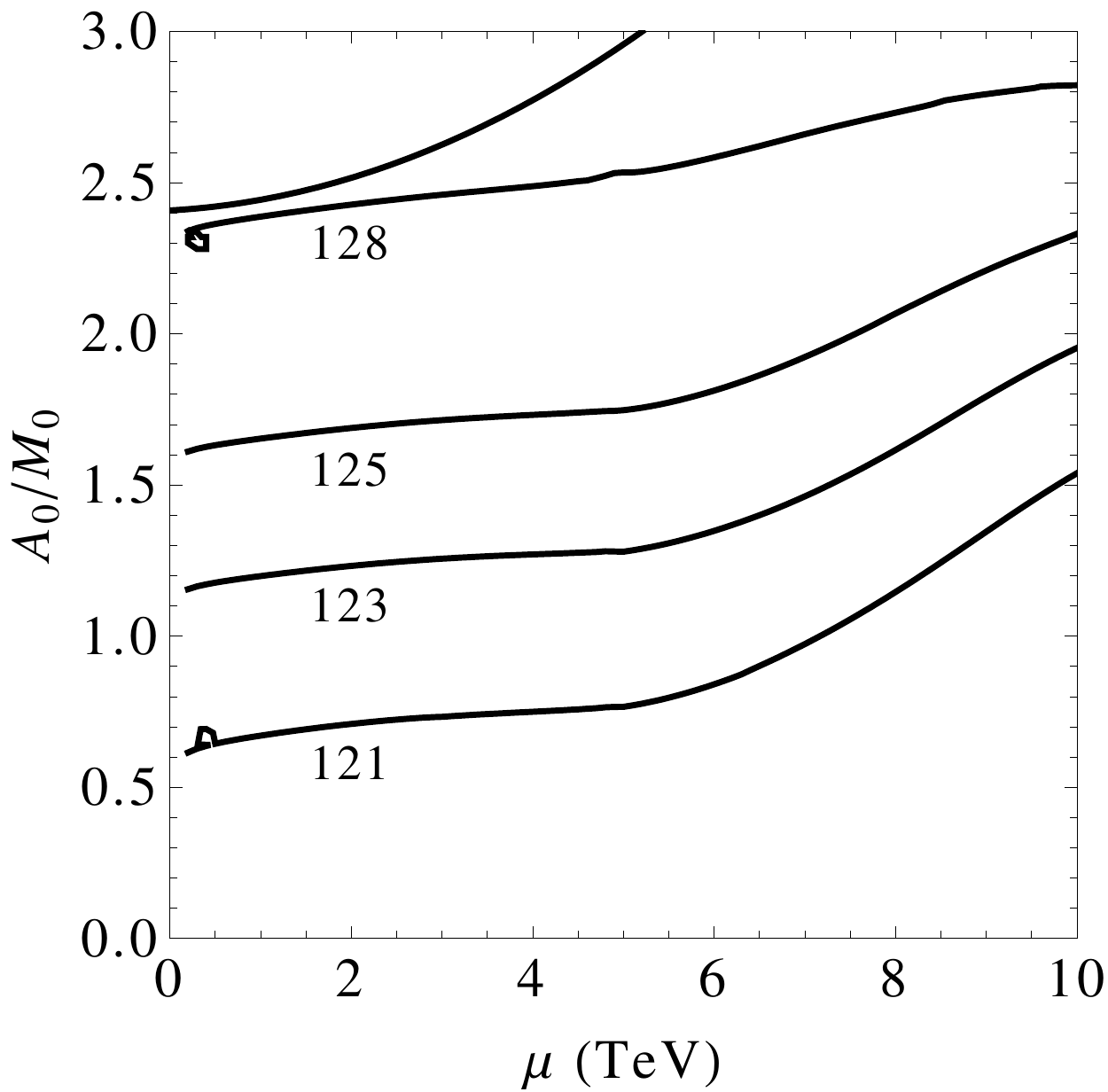}
\caption{For each combination of $M_0$, $m_A$, $M_1$, and $\tan\beta$, we use the relic density and Higgs mass constraints to fix the values of $A_0$ and $\mu$. The left and right panels show the relic density and Higgs mass, respectively, for an $A$-funnel model defined by $m_A=500$ GeV, $M_1=250$ GeV, $\tan \beta=20$, and $M_0=5$ TeV. The upper left regions are ruled out by vacuum stability constraints.} 
\label{fig:mu-A0}
\end{figure*}

\begin{figure*}[!]
\includegraphics[width=3.5in]{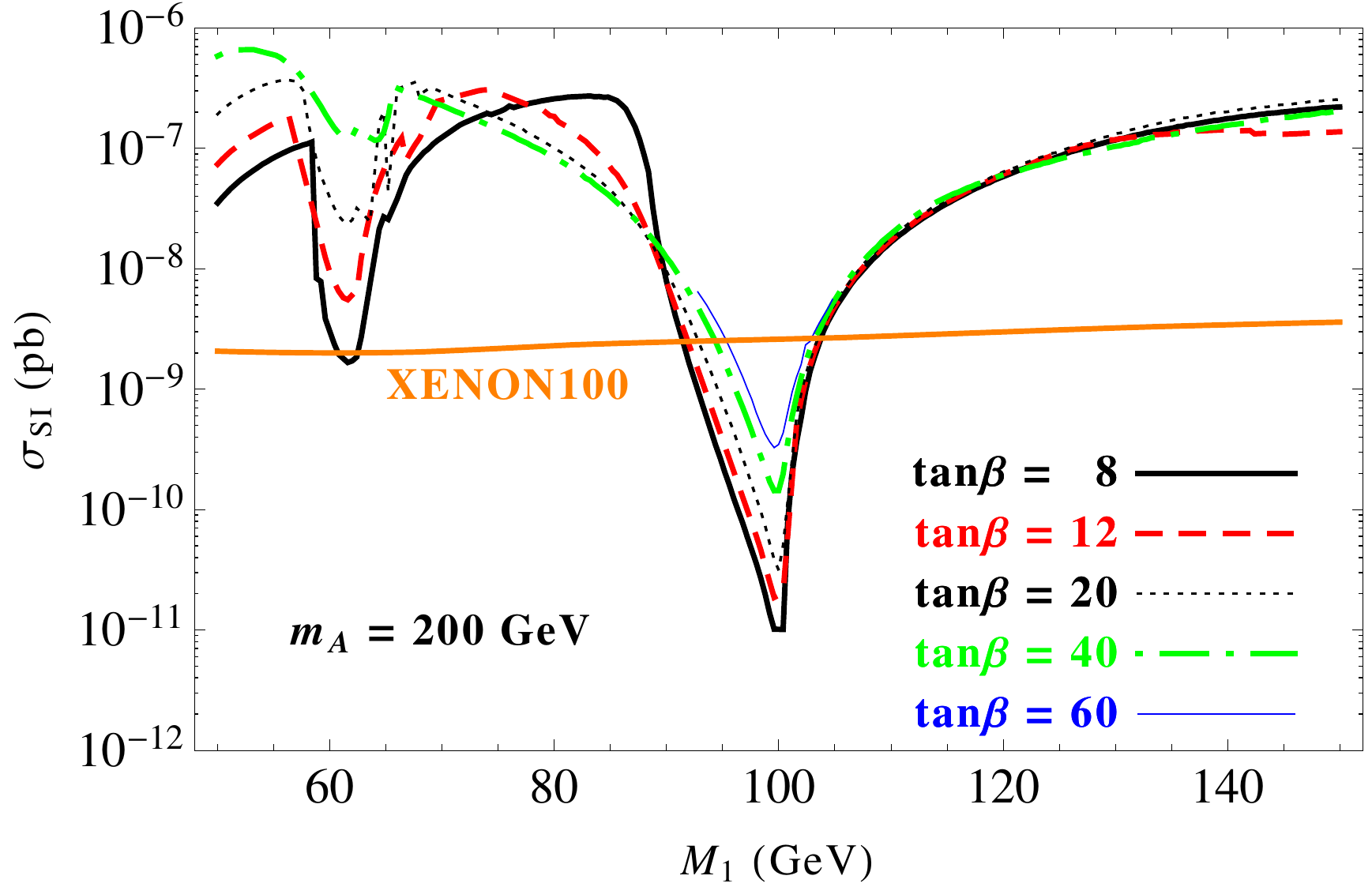}
\includegraphics[width=3.5in]{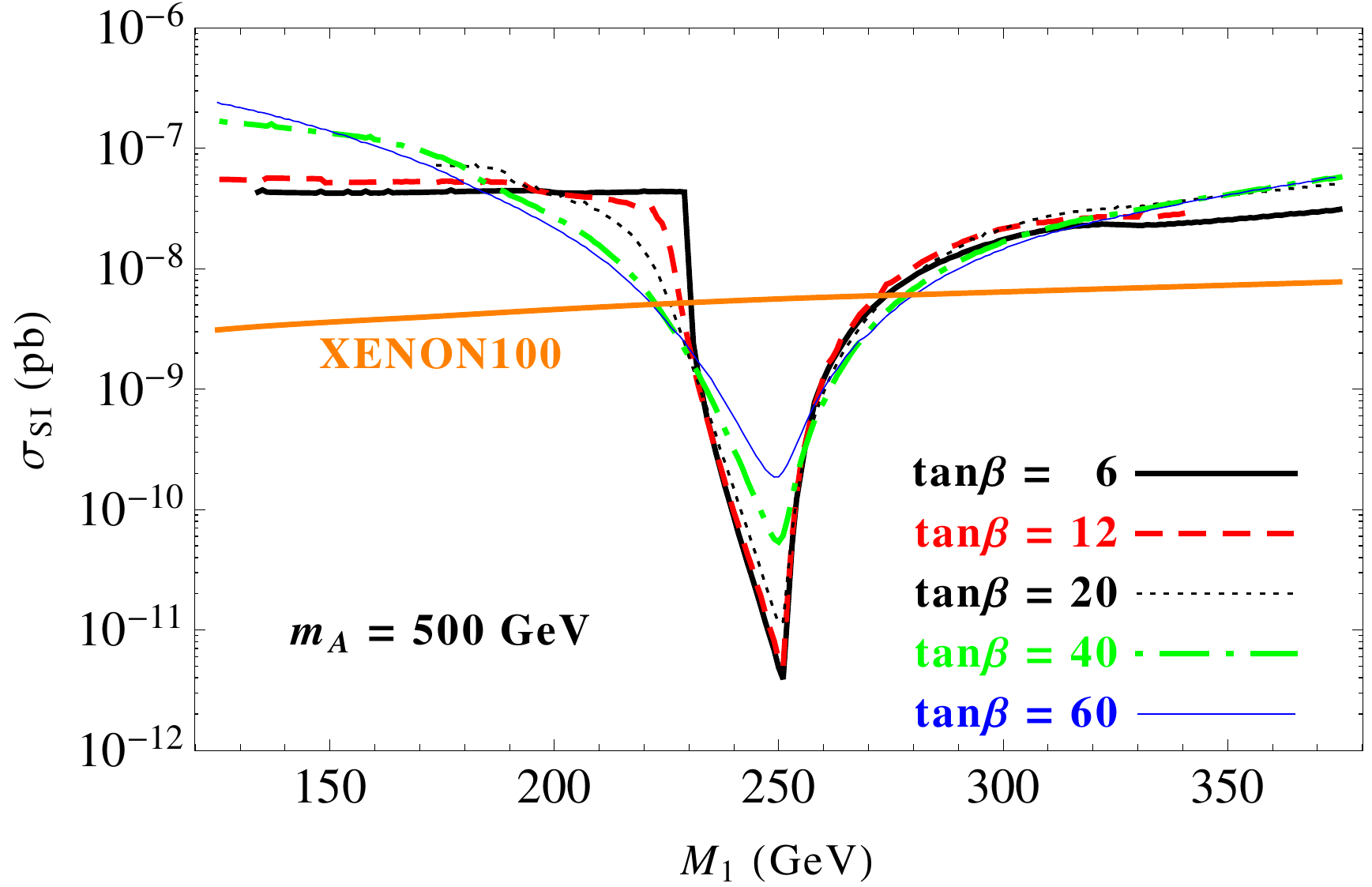}\\
\includegraphics[width=3.5in]{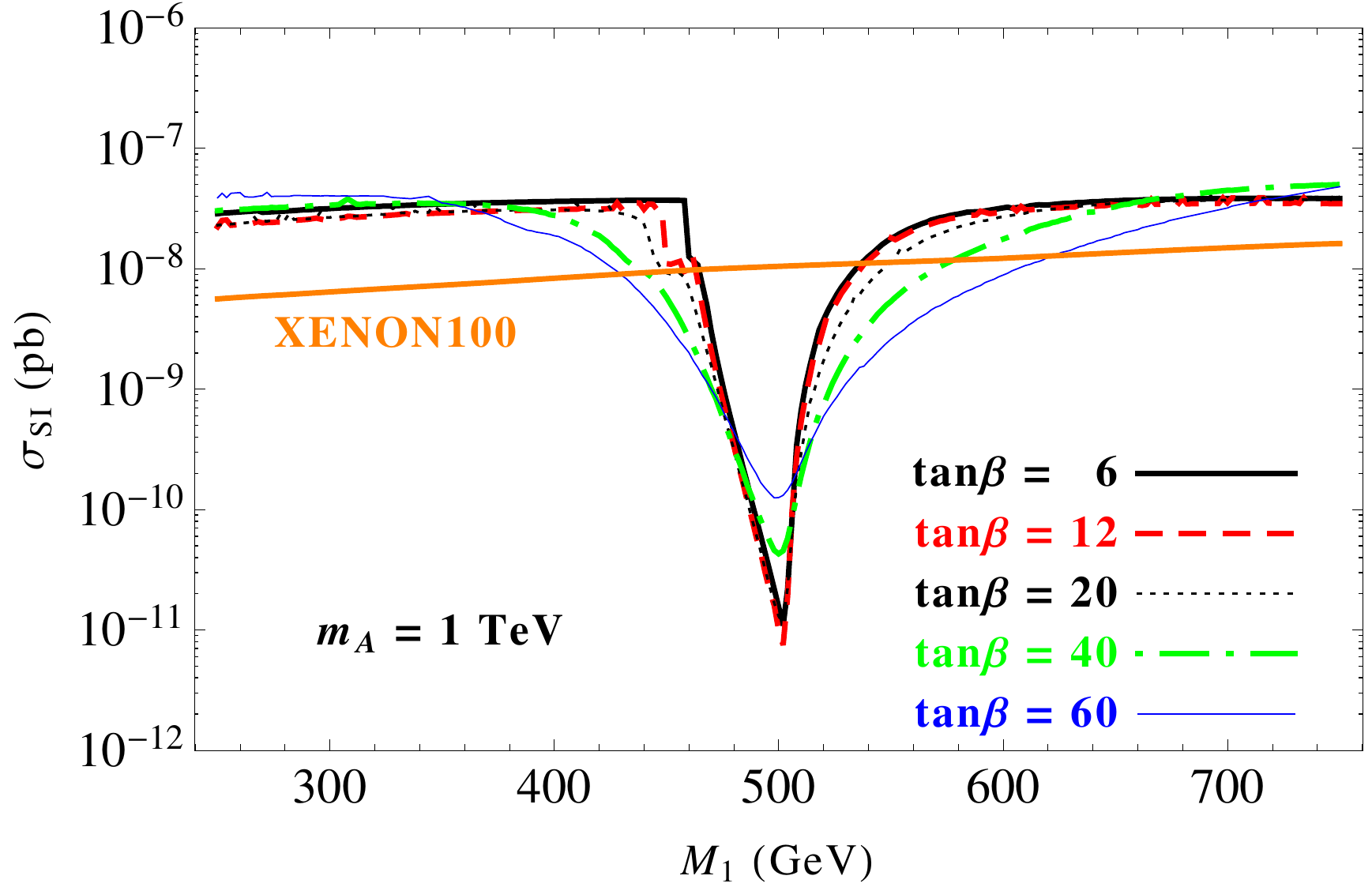}
\includegraphics[width=3.5in]{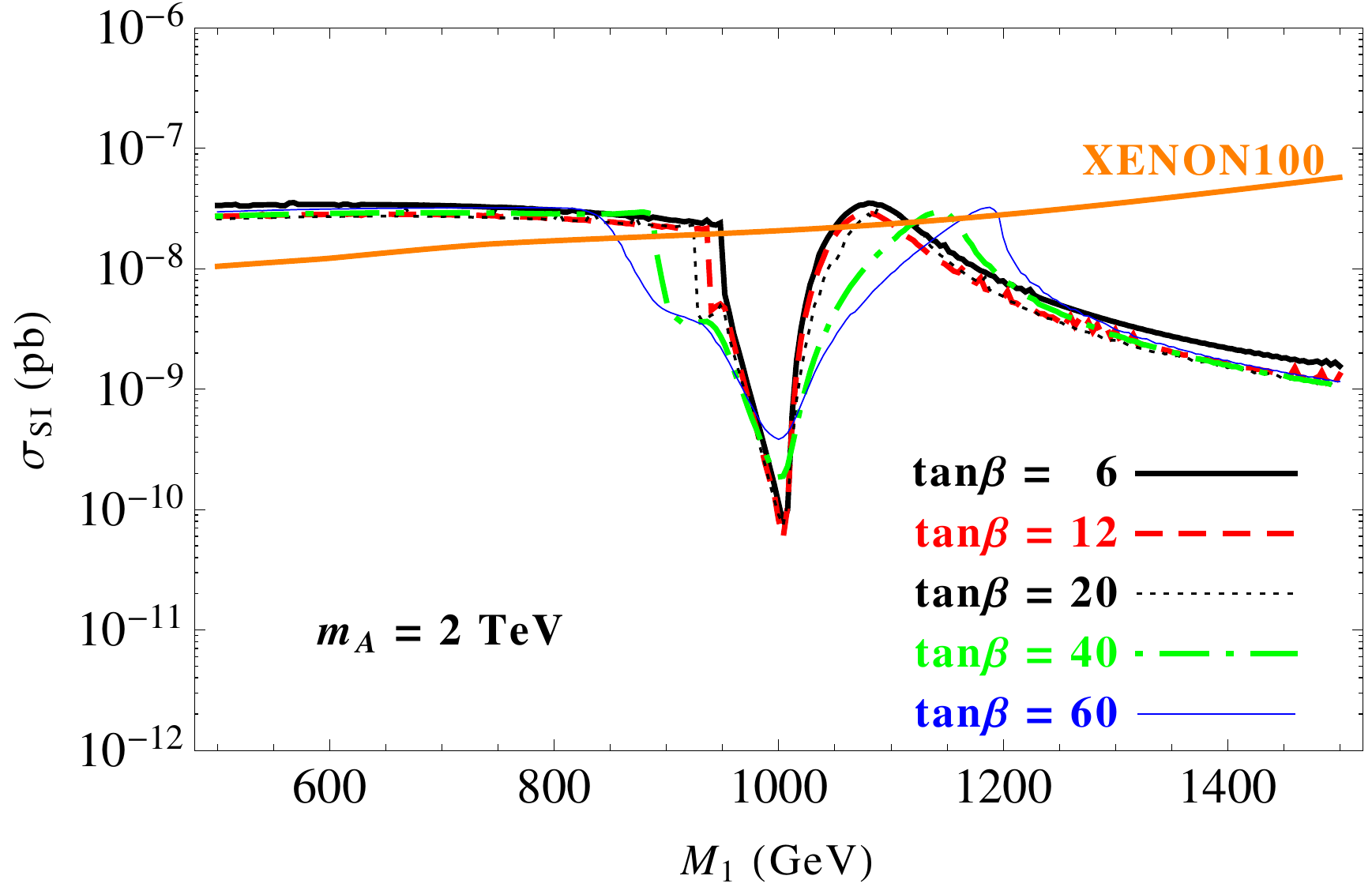}
\caption{The neutralino spin-independent elastic scattering cross section with nucleons as a function of $M_1$ for several values of $m_A$ and $\tan \beta$. We have set $M_2$, $M_3$ and the sfermion masses to 5 TeV and fixed $\mu$ and $A_0$ to obtain the measured dark matter density and Higgs mass. Also shown are the current constraints from XENON100, plotted under the approximation that $m_{\chi^0}=M_1$.} 
\label{fig:tanBeta}
\end{figure*}


\section{Neutralino Dark Matter in the $A$-Funnel}
\label{Afun}

The tree level diagrams for neutralino-quark spin-independent elastic scattering include $t$-channel CP-even Higgs exchange and $s$-channel squark exchange. In light of stringent constraints on squark and gluino masses from the LHC~\cite{squarksgluinos}, and in order to focus on the $A$-funnel region, we set all sfermion and gaugino masses (except $M_1$) to a high mass scale, $M_0$.\footnote{Note that our results do not change significantly if we instead adopt a hierarchy of gaugino masses that evolve to a common value at a high scale.}
%
After this simplifying set of assumptions, the following parameter set completely defines our MSSM:  
\[\left(M_0,m_A,\tan\beta,M_1,\mu,A_0\right).\]

Here, $m_A$ is the mass of the pseudoscalar Higgs, $\tan \beta$ is the ratio of the vacuum expectation values of the two Higgs doublets, $M_1$ is the bino mass, $\mu$ is the higgsino mass parameter, and $A_0$ is the common trilinear coupling.

We begin by imposing two constraints: that the thermal relic density of neutralinos (as calculated using MicrOmegas~\cite{Belanger:2010pz}) falls within the dark matter abundance as reported by WMAP~\cite{Hinshaw:2012fq}, and that the mass of the lightest Higgs boson (as calculated using FeynHiggs~\cite{Heinemeyer:2007aq}) falls within the range reported by ATLAS and CMS (which we generously take to be 123-128 GeV). In practice, we use this pair of constraints to set the values of $\mu$ and $A_0$. In Fig.~\ref{fig:mu-A0}, we show the relic density and Higgs mass as a function of $\mu$ and $A_0$ for the case of $M_0=5\,$TeV, $m_A=500\,$GeV, $M_1=250\,$GeV, and $\tan\beta=20$. From these figures, it is clear that these constraints are nearly orthogonal on the $A_0$-$\mu$ parameter space, making it possible to fix both of these values. 


With the values of $\mu$ and $A_0$ set (for any given combination of $M_0$, $m_A$, $M_1$ and $\tan\beta$), we proceed to evaluate the elastic scattering cross section of the lightest neutralino and compare this to constraints from XENON100. In Fig.~\ref{fig:tanBeta}, we plot this cross section as a function of $M_1$ (which in most cases shown is roughly equal to the mass of the lightest neutralino), for several values of $m_A$ and $\tan \beta$.

\begin{figure*}[t!]

\includegraphics[width=3.0in]{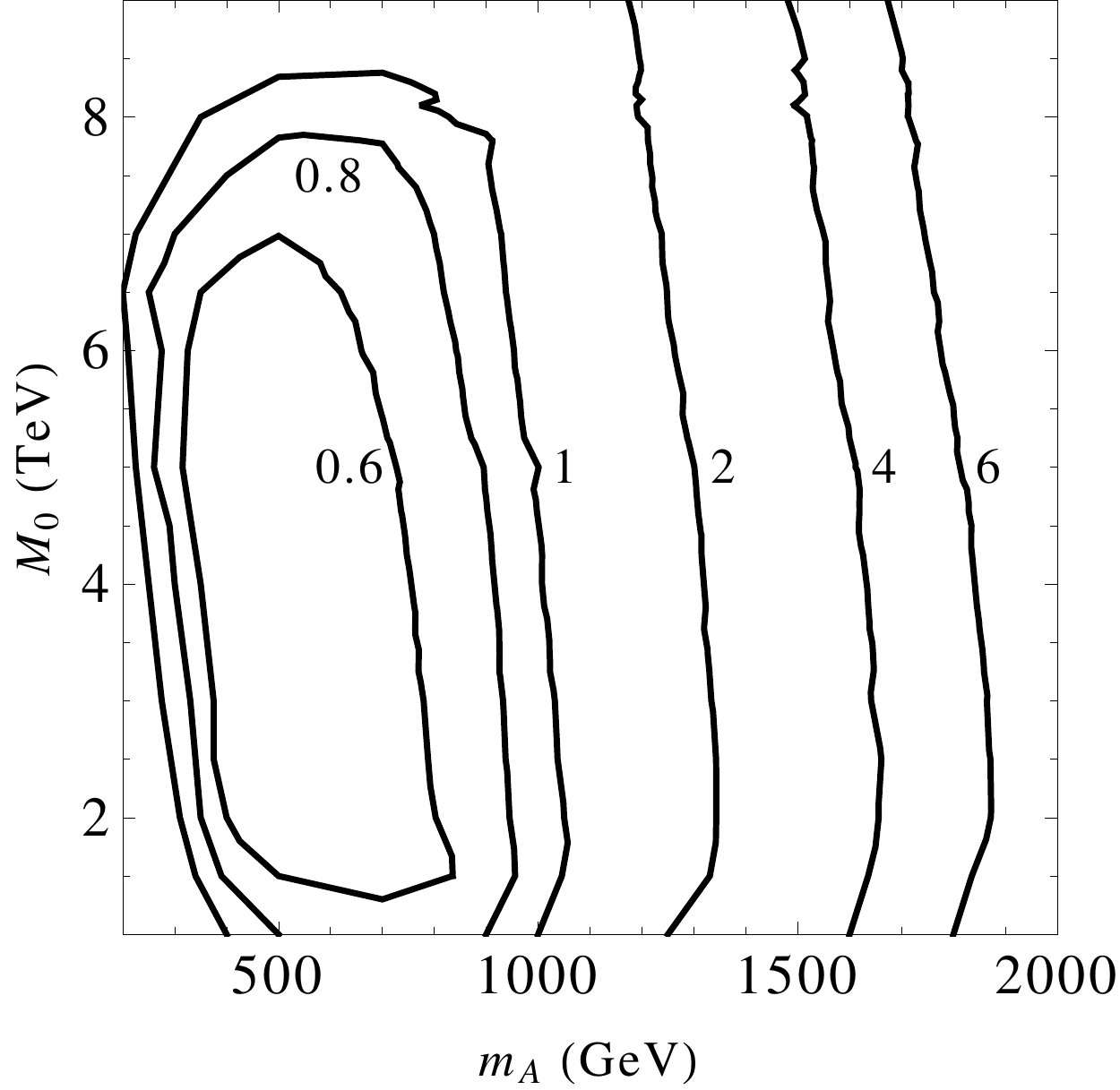}
\includegraphics[width=3.0in]{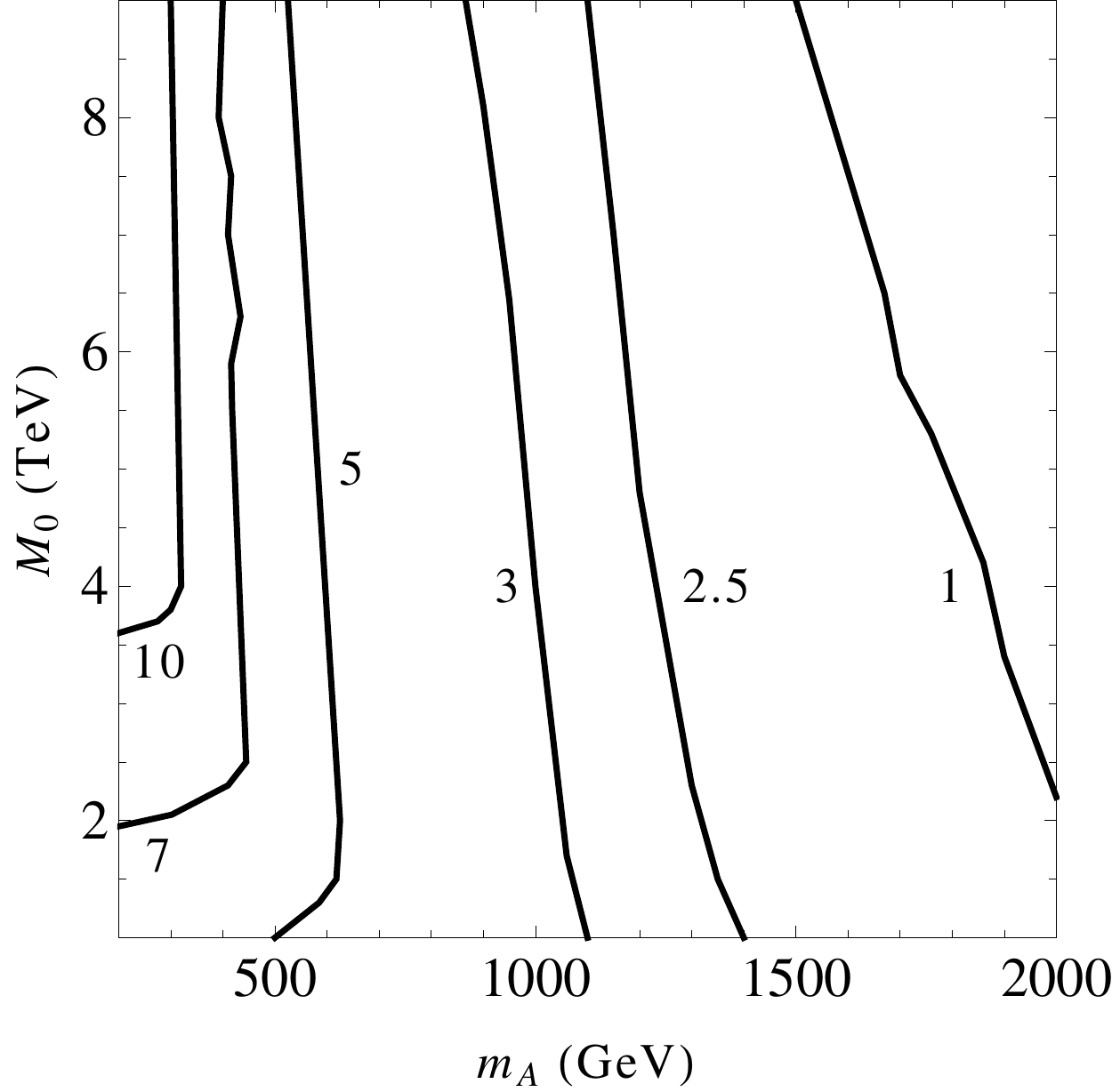}\\
\caption{In the left frame, we plot the minimum elastic scattering cross section with nucleons (in units of $10^{-11}$ pb) as a function of $M_0$ and $m_A$. In the right frame, we show the value of $\mu$ (in TeV) which yields the minimum elastic scattering cross section. Each of these points predict a relic density within WMAP's $2\sigma$ range and a Higgs mass within the appropriate range (123-128 GeV). } 
\label{fig:m0-mA}
\end{figure*}

\pagebreak

{\tiny.}

{\tiny.}

As anticipated, the XENON100 constraint rules out any value of $M_1$ up to $\sim$1 TeV that is not fairly close to the $A$ resonance ($m_A \approx 2 M_1$). For values near the resonance, however, these direct detection constraints can be evaded. Note that we only plot results for parameter values that yield acceptable relic densities and Higgs masses. For example, there are no values at all for $\tan\beta=4$, while the thick black line for $\tan\beta=6$ extends only a short distance up the funnel (for small $\tan\beta$, the Higgs mass can not be made large enough without introducing problems with vacuum stability). Note that the $\tan\beta=40$ and 60 contours in the $m_A=200$ GeV frame have been excluded by the LHC~\cite{higgstoditau}, and should be taken only as a point of comparison.




To ensure that the results shown in Fig.~\ref{fig:tanBeta} do not depend strongly on our (somewhat arbitrary) choice of $M_0$, we consider the impact of this parameter in the left frame of Fig.~\ref{fig:m0-mA}. As can be seen, while the depth of the $A$-funnel (the minimum elastic scattering cross section) depends mildly on the value of $M_0$ that is being considered, this variation is quite small compared to the much larger range shown in Fig.~\ref{fig:tanBeta}. Throughout the entire range of $M_0$ and $m_A$, the minimum cross section for $A$-funnel models is always within a factor of a few of $10^{-11}$ pb. 

For a viable thermal neutralino model to predict such a small elastic scattering cross section, the parameters $\mu$, $A_0$, and $\tan\beta$ must each take on rather specific values. First of all, at the very bottom of the funnel, annihilations are very efficient, and $|\mu|$ must be very large (and the neutralino's couplings to the $A$ must be very small) to accommodate the measured dark matter abundance, as evident in the right panel of Fig.~\ref{fig:m0-mA}. As noted earlier, we also require large values of $A_0$ (typically 1.3-2.5\,$M_0$) to accommodate the Higgs mass measured at the LHC. The maximum depth of the funnel occurs for intermediate values of $\tan \beta$, typically around $\sim$$10$.

Lastly, we note that we have not explicitly scanned over negative values of $\mu$ in this study.  In the case of negative $\mu$, loop-level cancellations can in some cases enhance the $A$ and $H$ decay widths~\cite{Altmannshofer:2012ks}, potentially making the $A$-funnel more broad and less deep. Furthermore, the coupling of the lightest neutralino to the light Higgs can be suppressed relative to the positive case. In the parameter space with small or moderate $\tan\beta$ and high $m_A$ (where the elastic scattering is dominated by light Higgs exchange), the choice of negative $\mu$ can lead to direct detection rates that are smaller by a factor of a few than those presented here~\cite{Feng:2010ef}.


\section{Projections: Toward Closing the Funnel}
\label{projections}

When direct detection experiments become sensitive to dark matter-nucleon cross sections as small as $\sigma \sim 3\times 10^{-12}$ pb, the $A$-funnel region of supersymmetric parameter space will be ruled out. Current constraints are three orders of magnitude away from this sensitivity, however, and it will most likely be another decade or more before such tiny cross sections will be probed. Fortunately, it is unlikely that nature will be described by a dark matter model that lies at the absolute minimum of such a resonance. In this section, we will quantify how precisely one needs to tune such a resonance in order for neutralino dark matter to provide the measured thermal abundance without violating current and projected direct detection constraints.

At present, in order for an $A$-funnel model to be viable, the neutralino mass must be fine tuned to a modest degree, not far from $m_{\chi} \approx m_A/2$. From Fig.~\ref{fig:tanBeta}, we can see that the neutralino mass must fall within roughly 10\% of $m_A/2$ if direct detection constraints are to be evaded. And while this is not yet a highly unacceptable degree of tuning, this requirement will become increasingly severe as the constraints placed by direct detection experiments become more stringent.

\begin{figure}[!t]
\centering
\includegraphics[width=0.95\columnwidth]{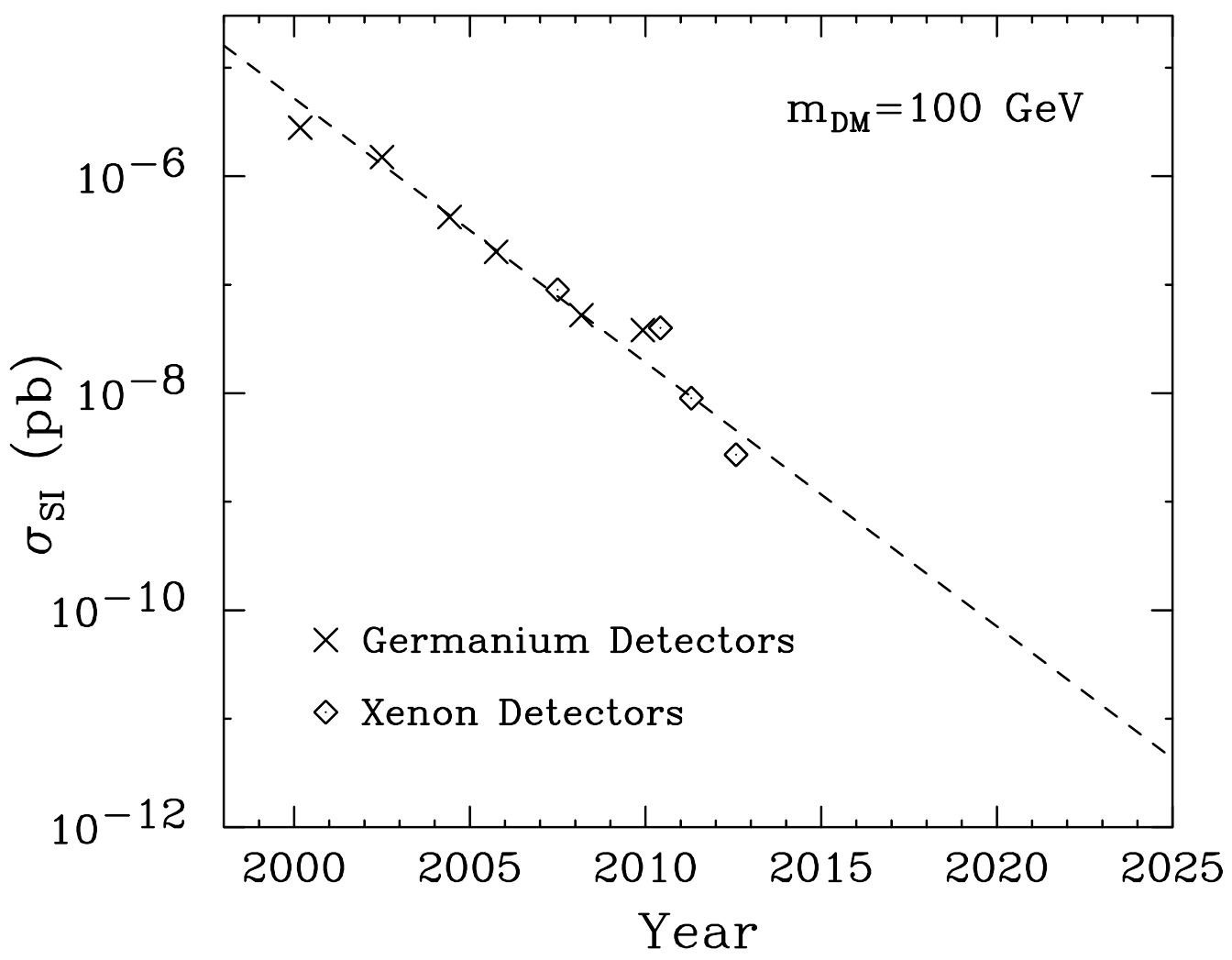}
\caption{Over the past decade, the sensitivity of direct detection experiments has improved with a Moore's Law-like behavior, strengthening their constraints by a factor of two every 15 months (the doubling time for Moore's law is actually 18 months). Shown here is the strongest constraint on the spin-independent WIMP-nucleon elastic scattering cross section, as a function of time, for a 100 GeV WIMP mass. In performing our projections, we assume that this simple extrapolation continues until reaching cross sections at the level of $\sim$$10^{-12}$ pb, where irreducible neutrino backgrounds are predicted to become important~\cite{bg}.}
\label{limits}
\end{figure}

Over the past decade, the sensitivities of direct detection experiments have collectively improved exponentially with time, strengthening their constraints on average by a factor of two every 15 months. For comparison, Moore's law describing the rate of advancement in computational performance exhibits a similar but somewhat slower doubling time of approximately 18 months. In Fig.~\ref{limits} we show the current and past constraints placed by a number of direct detection experiments (from the CDMS, Edelweiss, XENON10, and XENON100 collaborations) as a function of time, for the case of a 100 GeV WIMP mass~\cite{xenon,cdms,limitplotter}. The dashed line represents the exponential progress that has taken place over the past 13 years. 

There are compelling reasons to anticipate this rate of progress to continue, at least for the next several years. In the near term, LUX is anticipated to produce its first constraint by the end of 2013, and to reach a sensitivity of a few $\times 10^{-10}$ pb after roughly a year of live time~\cite{Akerib:2012ak}. In the meantime, construction of XENON1T is scheduled to begin later this year, with a projected reach of a few $\times 10^{-11}$ pb~\cite{Aprile:2012zx}. Even with a reasonable allotment for unanticipated delays, the prospects for meeting or exceeding the historical rate of progress appear very encouraging.\footnote{An argument could be made that in the transition from germanium-based to xenon-based detectors, the rate at which direct detection experiments are improving has increased, perhaps to a doubling time in sensitivity that is as short as 10 months. As encouraging as this may be, we chose to adopt the more conservative projection shown in Fig.~\ref{limits}.}

\begin{figure*}[htp]
\centering
\includegraphics[width=0.95\columnwidth]{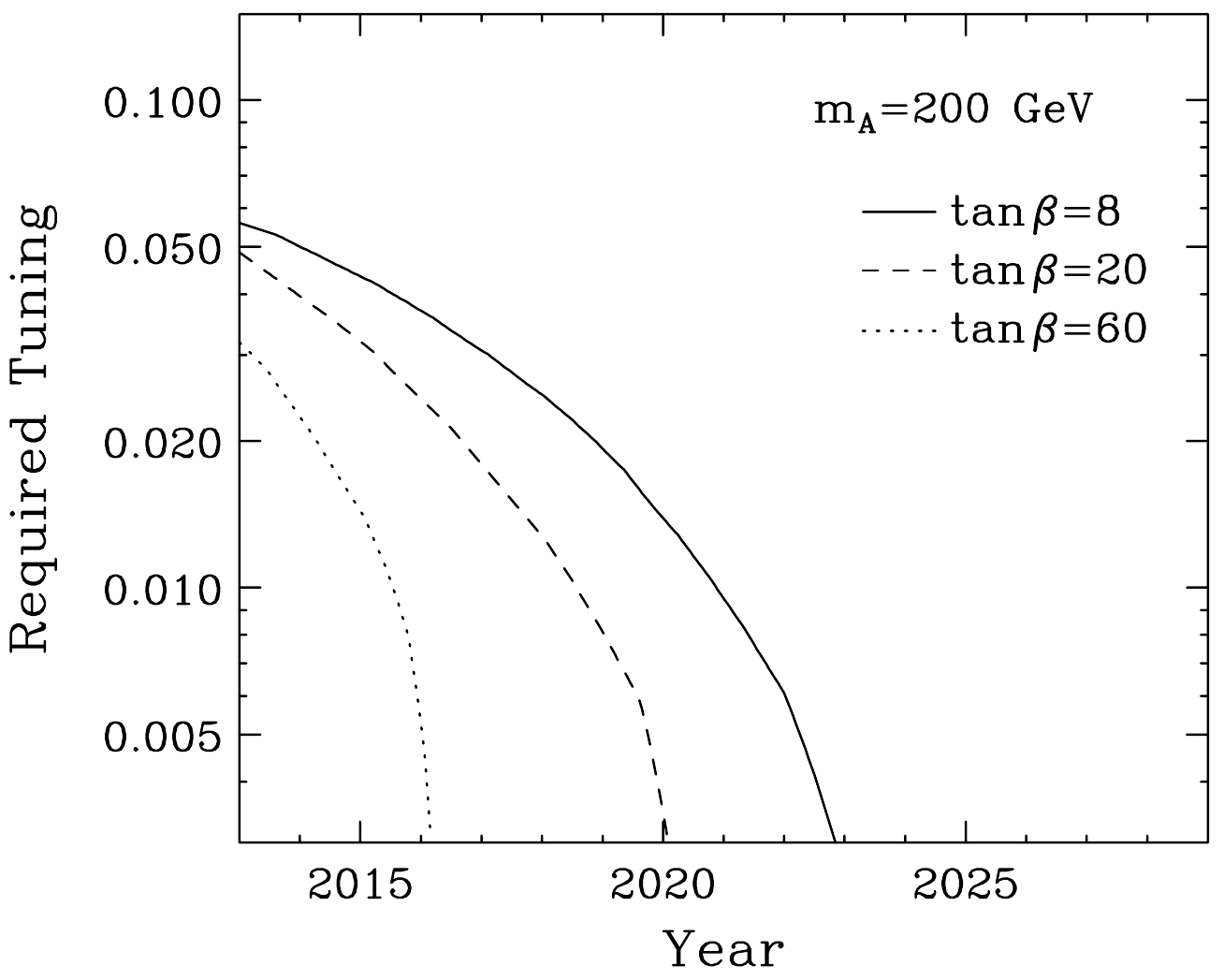}
\hspace{0.5cm}
\includegraphics[width=0.95\columnwidth]{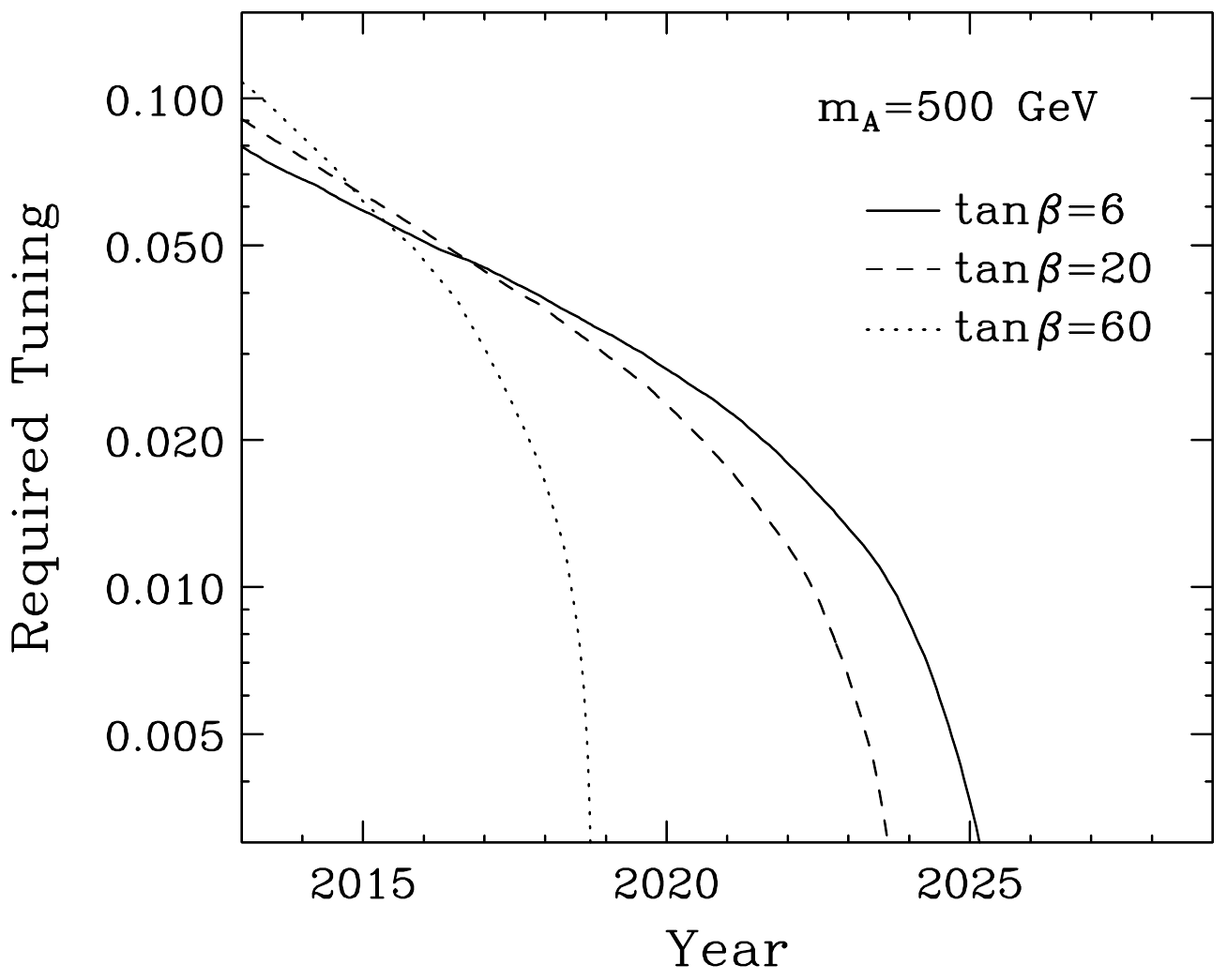}\\
\includegraphics[width=0.95\columnwidth]{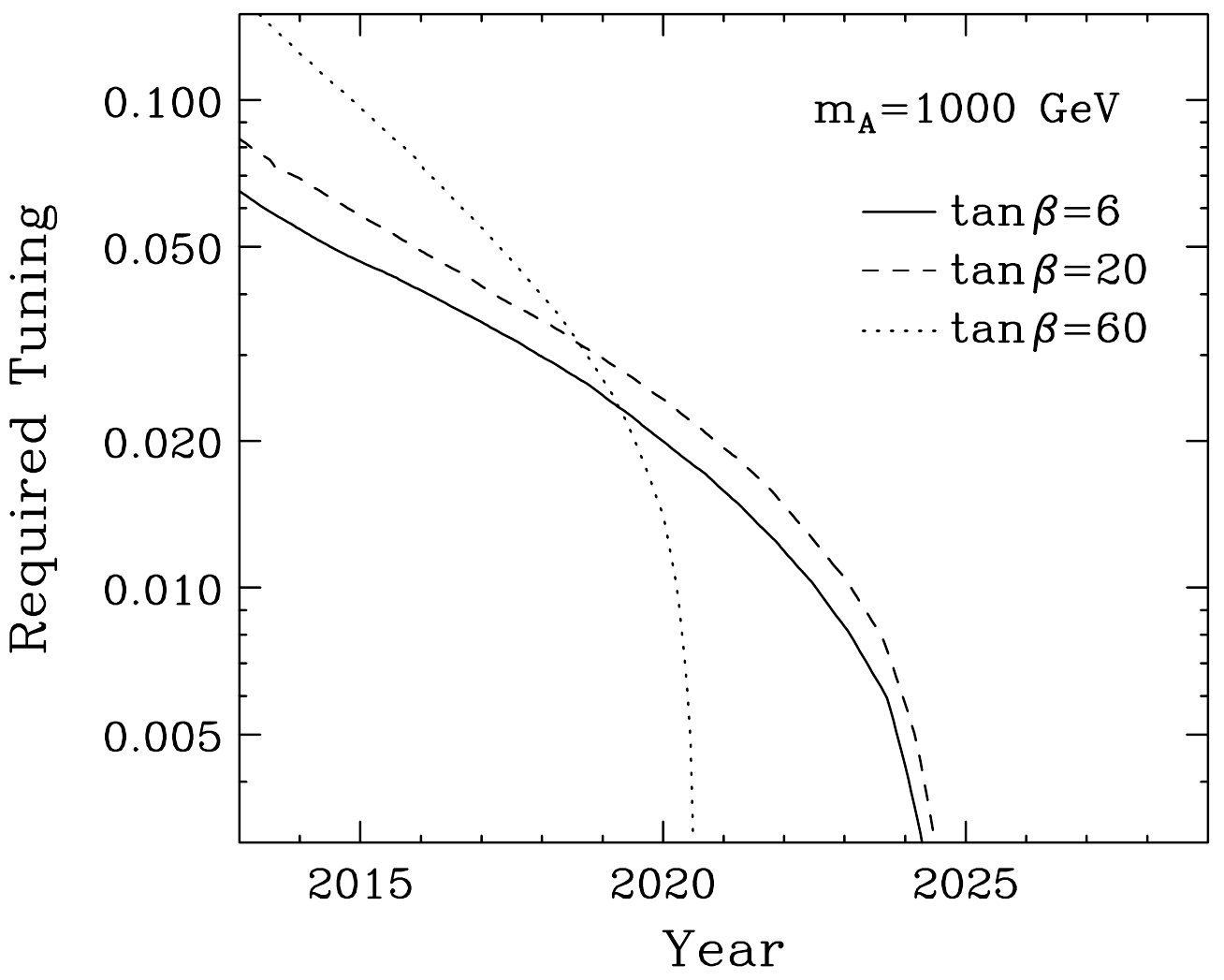}
\hspace{0.5cm}
\includegraphics[width=0.95\columnwidth]{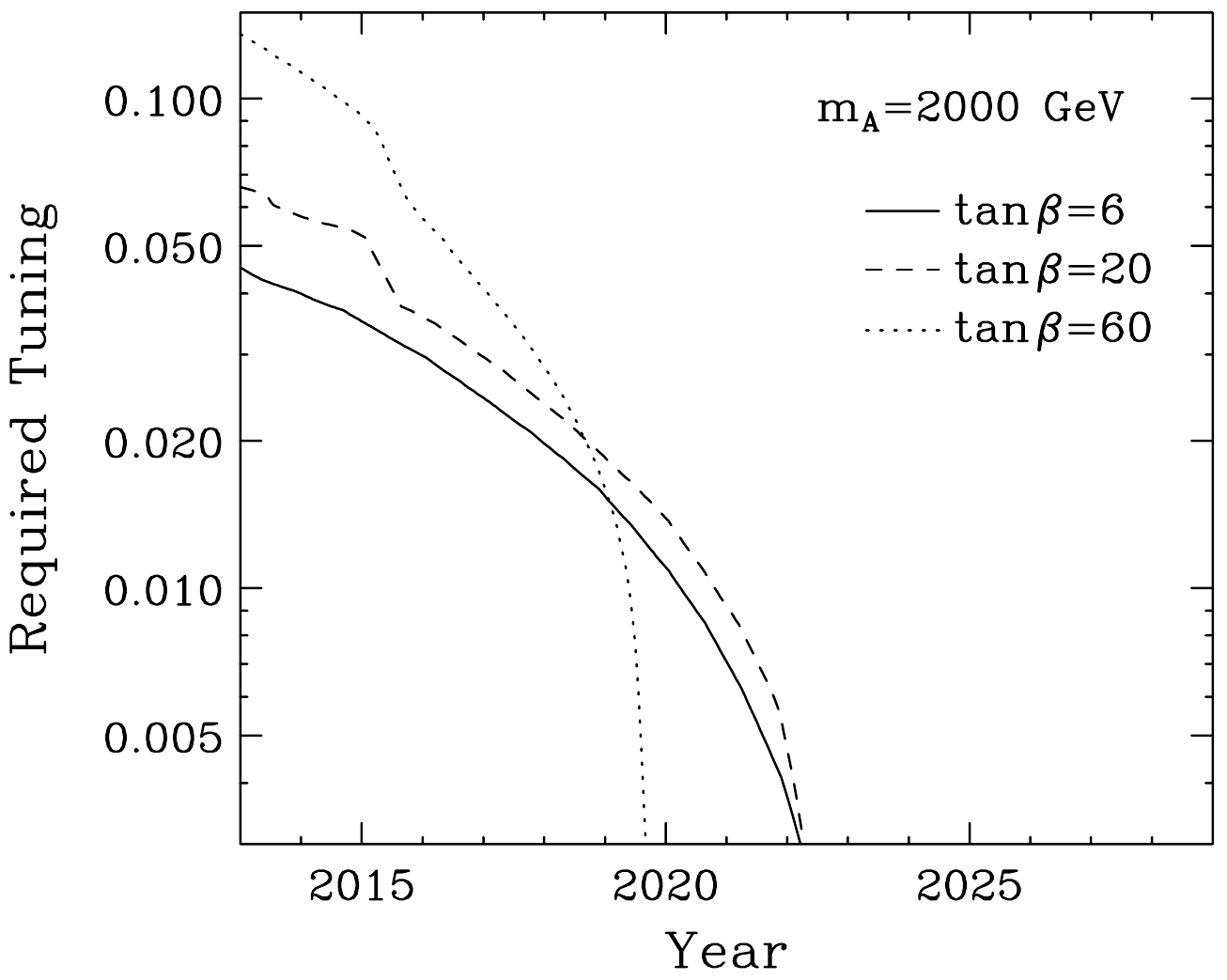}
\caption{The degree to which the neutralino's mass must be tuned to avoid current and projected constraints from direct detection experiments. To avoid current constraints, $m_{\chi^0}$ must lie within 3--16\% of the central value of the resonance.  Assuming the rate of progress shown in Fig.~\ref{limits}, in only five years 1--4\% tuning will be required. In approximately a decade, direct detection experiments are anticipated to reach the sensitivity needed to test even the most highly tuned resonances, closing the entire $A$-funnel region of parameter space.}
\label{tuning}
\end{figure*}

In Fig.~\ref{tuning}, we show how these projected improvements in direct detection sensitivity will impact the degree of fine tuning required for neutralinos in the $A$-funnel region of parameter space. With present constraints, $m_A$ must fall within 3--16\% of the central resonant value, depending on the values of $\tan \beta$ and $m_{\chi}$. Assuming the rate of progress shown in Fig.~\ref{limits} (and assuming that no detection is made), in only five years approximately 1-4\% tuning will be required. In a little over a decade, these experiments are anticipated to reach the sensitivity needed to test even the most highly tuned resonances, closing the entire $A$-funnel region of parameter space.

\section{Other Resonant Regions and Model Independent Considerations}
\label{others}

The pseudoscalar Higgs is not the only resonance through which neutralinos could efficiently annihilate. In this section, we will consider dark matter particles that annihilate through other resonances, including the light Higgs and $Z$ resonances of the MSSM. 

\subsection{The Light Higgs and $Z$ Poles}
\label{zhsec}

In the upper left frame of Fig.~\ref{fig:tanBeta}, in addition to the $A$-funnel, one can see the effect of the light Higgs resonance (at $M_1 \sim m_h/2$). While for the value of $m_A$ shown, the presence of the light Higgs resonance does not enable us to evade the constraints from XENON100, if we increase $m_A$ to a high value we can reduce the prediction for the elastic scattering cross section while still allowing for efficient annihilations through the light Higgs resonance~\cite{Han:2013gba}.

\begin{figure}[!t]
\centering
\includegraphics[width=0.99\columnwidth]{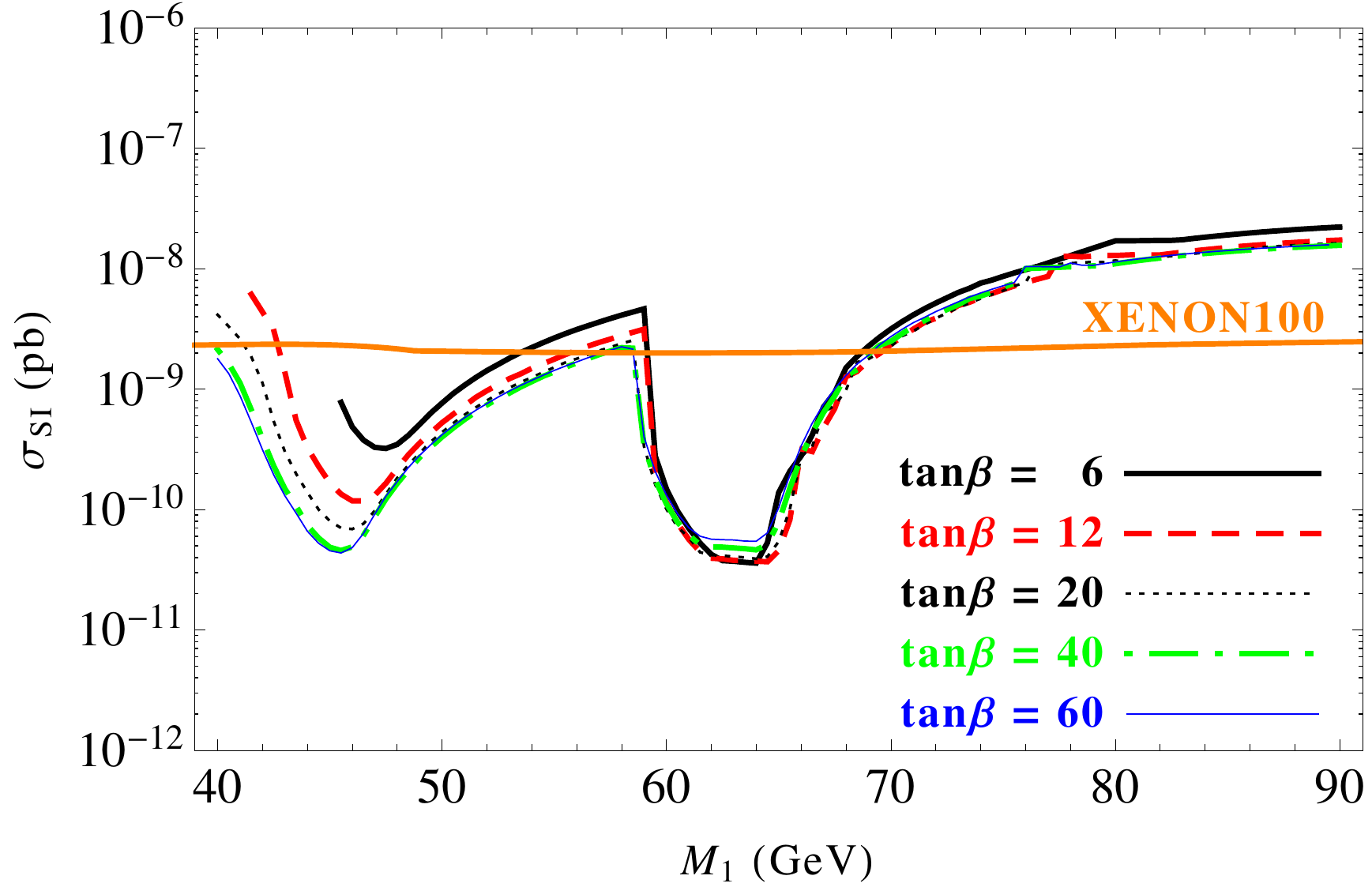}
\caption{The neutralino spin-independent elastic scattering cross section with nucleons as a function of $M_1$ for several values of $\tan \beta$, in the regions of the light Higgs and $Z$ resonances. We have set $M_2$, $M_3$, $m_A$ and the sfermion masses to 5 TeV and fixed $\mu$ and $A_0$ to obtain the measured dark matter density and Higgs mass.} 
\label{zh}
\end{figure}

In Fig.~\ref{zh} we focus on this possibility, along with that of neutralinos annihilating through the $Z$ resonance. Here, we have set $m_A$ to 5 TeV, but otherwise follow the same procedure as described in Sec.~\ref{Afun}. Notice that while these resonances are sufficient to potentially evade current direct detection constraints, they do not extend to elastic scattering cross sections that are as small as those found in the $A$-funnel. Based on the extrapolation shown in Fig.~\ref{limits}, we project that both the light Higgs and $Z$ resonance regions will be closed by direct detection experiments within the next approximately 7 years.

\subsection{Other (non-MSSM) Resonances}

Many of the arguments and conclusions discussed in this article apply to a broad class of scenarios in which the dark matter annihilates through an efficient resonance, supersymmetric or otherwise. In this subsection, we briefly comment on the current and future viability of resonant scenarios that might be found outside of the context of the MSSM.

For a general dark matter candidate that annihilates through a scalar resonance (such as through a scalar Higgs), the arguments presented here for the case of the MSSM remain valid, with conclusions that are essentially unchanged. Other types of particles that might be exchanged on resonance, however, can in some cases evade these considerations. For example, in a model in which the dark matter annihilates through a pseudoscalar (without the accompanying scalars found in the MSSM), no low-velocity elastic scattering cross section is generated. Similarly, if the dark matter is a Majorana fermion, it could annihilate through a vector resonance without generating a corresponding spin-independent elastic scattering cross section (although a spin-dependent interaction would be induced, the current constraints on spin-dependent scattering are much less stringent). The reason that the $Z$ pole region of the MSSM can be successfully probed by direct detection experiments (see Sec.~\ref{zhsec}) is that a neutralino near the $Z$ pole must have a significant bino component (due to LEP constraints on chargino masses) and thus will couple significantly to the light Higgs. A purely-higgsino neutralino near the $Z$ pole that somehow was not accompanied by a corresponding light chargino could satisfy all direct detection constraints, although IceCube would likely be able to rule out such a scenario~\cite{icecube}. More realistically, one could imagine a heavier vector resonance ({\it i.e.}~a $Z'$) though which a Majorana dark matter candidate could annihilate without violating direct detection constraints.

\section{Conclusions}
\label{conclusions}

In this article, we have discussed the viability of resonant regions of the MSSM, in which the lightest neutralino avoids being over produced in the early universe by annihilating through the $A/H$, $h$ or $Z$ resonance. At present, direct detection constraints can be respected if the neutralino's mass falls within about 16\% of the central resonant value (for some mass ranges and values of $\tan \beta$, the neutralino's mass is already constrained to fall even closer to the resonance, within as little as 3\% in some cases). This degree of tuning is comparable to the mass splittings required in viable coannihilation regions of the MSSM, and appears to be somewhat less severe than the degree of electroweak fine tuning required in currently viable regions of the MSSM~\cite{Feng:2013pwa}.

As direct dark matter detection experiments continue to increase in sensitivity, the degree to which the neutralino's mass will be required to fall near a resonance (or near the mass of a coannihilating superparticle) will increase. In particular, as such constraints become more stringent and requirements on the neutralino's couplings to Higgs bosons become more severe, overproduction in the early universe will occur unless the neutralino's mass is very close to the optimal value of the resonance. Assuming that direct detection experiments improve in sensitivity at a rate similar to that observed over the past decade, they will be able to entirely close the light Higgs and $Z$ resonance regions within the next approximately 7 years. At that time, the $A$-funnel region will remain viable, but will require that the neutralino mass (or equivalently, $m_A$) resides within 4\% of the central resonant value. At this rate of progress, direct detection experiments will reach the sensitivity required to close the $A$-funnel region of parameter space in approximately 12 years, even for optimal values of the masses.

\bigskip

{\it Acknowledgements}:  This work has been supported in part by the US Department of Energy. Support and resources from the Center For High Performance Computing at the University of Utah are gratefully acknowledged.

\end{document}